\documentclass[]{elsarticle}

\usepackage{hyperref}

\usepackage{natbib}
\usepackage{amssymb}

\journal{Planetary and Space Science}

\biboptions{authoryear}

\begin{document}

\begin{frontmatter}

\title{Resonances in the asteroid and trans--Neptunian belts: a brief review}

\author{Tabar\'e Gallardo}
\address{Instituto de F\'{i}sica, Facultad
	de Ciencias, UdelaR, Igu\'{a} 4225, 11400 Montevideo, Uruguay}


\cortext[mycorrespondingauthor]{Corresponding author}
\ead{gallardo@fisica.edu.uy}


\begin{abstract}
Mean motion resonances play a fundamental role in the dynamics of the small bodies of the Solar System. The last decades of the 20th century gave us a detailed description of the dynamics as well as the process of capture of small bodies in  coplanar or small inclination resonant orbits. More recently, semianalytical or numerical methods allowed us to explore the behavior of resonant motions for arbitrary inclination orbits. The emerging dynamics  is very rich, including large orbital changes due to secular effects inside mean motion resonances.
The process of capture in highly inclined or retrograde resonant orbits was  addressed showing that the capture in retrograde resonances is more efficient than in direct ones. 
A new terminology appeared in order to characterize the properties of the resonances.  
Numerical explorations in the transneptunian region showed the relevance and the particular dynamics of the exterior resonances with Neptune  which can account for some of the known high perihelion orbits in the scattered disk.
Moreover, several asteroids evolving in resonance with  planets other than  Jupiter or Neptune were found and a large number of asteroids in three-body resonances were identified. 
\end{abstract}


\end{frontmatter}

%\linenumbers

\section{Introduction}
\label{intro}

An orbital resonance occurs when there is a commensurability between frequencies associated with the orbital motion of some bodies. These frequencies can include the mean motion $n$ of the bodies (in which case we speak of a mean-motion resonance), or exclusively secular (low) frequencies associated with the  long term evolution of the longitude of the nodes, $\Omega$ or the longitude of the perihelia, $\varpi$. 
In the dynamics of small Solar System bodies, these commensurabilities can generate two-body mean-motion resonances, involving
the mean longitudes of the asteroid and one planet, three-body mean motion resonances,  involving the mean longitudes of the asteroid and two planets, secular resonances involving longitudes of the perihelia and nodes and the Kozai-Lidov (KL) mechanism involving the asteroid's argument of the perihelion, $\omega=\varpi-\Omega$ \citep{2017ASSL..441.....S}. A very concise but complete review on orbital resonances can be found in \citet{Malhotra98orbitalresonances}.
In this paper we will refer only to two-body mean motion resonances (hereafter 2BRs) and three-body mean motion resonances (hereafter 3BRs) or, in general, mean motion resonances (hereafter MMRs). 
We will focus on the main advances of the 21st century, for earlier reviews the reader may consult for example \citet{2002aste.conf..379N}, \citet{Malhotra98orbitalresonances} or \citet{1976ARA&A..14..215P}.

When an asteroid, or more generally, a minor body is in a 2BR with a planet of mass $m_1$
their mean motions verify
 $k_0n_0 + k_1n_1 \sim 0$ being $n_0$ and $n_1$ the mean motions of the minor body and the planet respectively and $k_0$ and $k_1$ small integers  with different sign.
In that case we say that the asteroid is in the resonance $|k_1|:|k_0|$.
From theories developed and valid for low-inclination orbits it was proved that the resonance's strength  is approximately proportional to $m_1e^q$, being $e$ the orbital eccentricity of the resonant minor body and where $q=|k_0+k_1|$ is the order of the resonance \citep{1999ssd..book.....M}. It turns out that when considering low-inclination orbits,  being $e<1$, only low order resonances have dynamical interest (the high-order ones have negligible strength). The above criteria for resonant motion is just an approximation and the precise definition of the resonant state is given by the behavior of the critical angle $\sigma = k_0\lambda_0 + k_1\lambda_1 + \gamma$ being $\lambda_i$ the quick varying mean longitudes and $\gamma$ a slow evolving angle defined by a linear combination of the $\Omega_i$ and $\varpi_i$ involved.
A resonant motion is characterized by an oscillation, or \textit{libration}, of the critical angle around a stable equilibrium point. In the low-inclination approximation they are located at   $\sigma=0^{\circ}$ or  $\sigma=180^{\circ}$
except for exterior resonances of the type 1:k and 1:1 resonances for which  the locations depend on the orbital eccentricity, which is why they are known as asymmetric. A very special case of 2BR that has deserved a lot of attention along the history of celestial mechanics since Lagrange's times is the strong 1:1 resonance, that means coorbital objects like Jupiter's trojans and  quasi-satellites.

On the other hand a minor body  is in a 3BR with two planets of mass $m_1$ and $m_2$ when the mean motions  verify $k_0n_0 + k_1n_1 +k_2n_2 \sim 0$. From  theories developed for zero inclination orbits it was proved that the resonance's strength  is approximately proportional to $m_1m_2e^q$, where $q=|k_0+k_1+k_2|$ is the order of the resonance \citep{1999CeMDA..71..243N}. It is clear that being the masses expressed in units of solar masses the 3BRs are orders of magnitude weaker than 2BRs. It is important to stress that the 3BRs are not necessarily the result of the superposition of 2BRs between the intervening bodies as can be the case of the Galilean satellites of Jupiter or some extrasolar planetary systems \citep{2016Icar..274...83G}. Three-body resonances exhibit also asymmetric equilibrium points as was showed by \citet{2014Icar..231..273G}. 

The commensurabilities above mentioned generate, in the long term, mean planetary perturbations on the minor body that are very different from the perturbations that a non resonant minor body experiences.
The small planetary perturbation given at the right frequency gradually sums up instead of canceling out.
 Resonances do not emerge as instantaneous dynamical effects as, for example, a close
encounter with a planet does. On the contrary, it is necessary to let the system evolve for several orbital revolutions in order that the minor body starts to feel the resonant gravitational potential.

The resonant motion is characterized by a regular small amplitude oscillation of the semi-major axis which preserves its mean value constant over time. This mean value is given
 by the corresponding mean motion $n_0$ defined by the resonant relation. These oscillations  are correlated with oscillations in the orbital eccentricity and the librations of the critical angle $\sigma$.
The frequency of the small amplitude oscillations are  related to the resonance's strength: stronger resonances exhibit higher frequency oscillations \citep{2007ASSL..345.....F}. These oscillations are a protective mechanism  that guarantees the constancy of the semimajor axis in front of other perturbations that the object can be exposed to. 
In particular, a chaotic diffusion of semi-major axis is immediately stopped (at least temporarily) if a capture in MMR occurs. This process is very common when simulating the orbital dynamics of minor bodies.
However, the long-term orbital evolution of the body inside the resonance can lead to major orbital changes in eccentricity and inclination, eventually bringing the small body close to collisions with other planets or the Sun.
 These large orbital variations are generated by secular mechanisms inside the resonance and not by the resonance itself. The secular mechanisms can be secular resonances, secondary resonances or the KL mechanism \citep{2002mcma.book.....M}.

\begin{figure}[h]
	\resizebox{12cm}{!}{\includegraphics{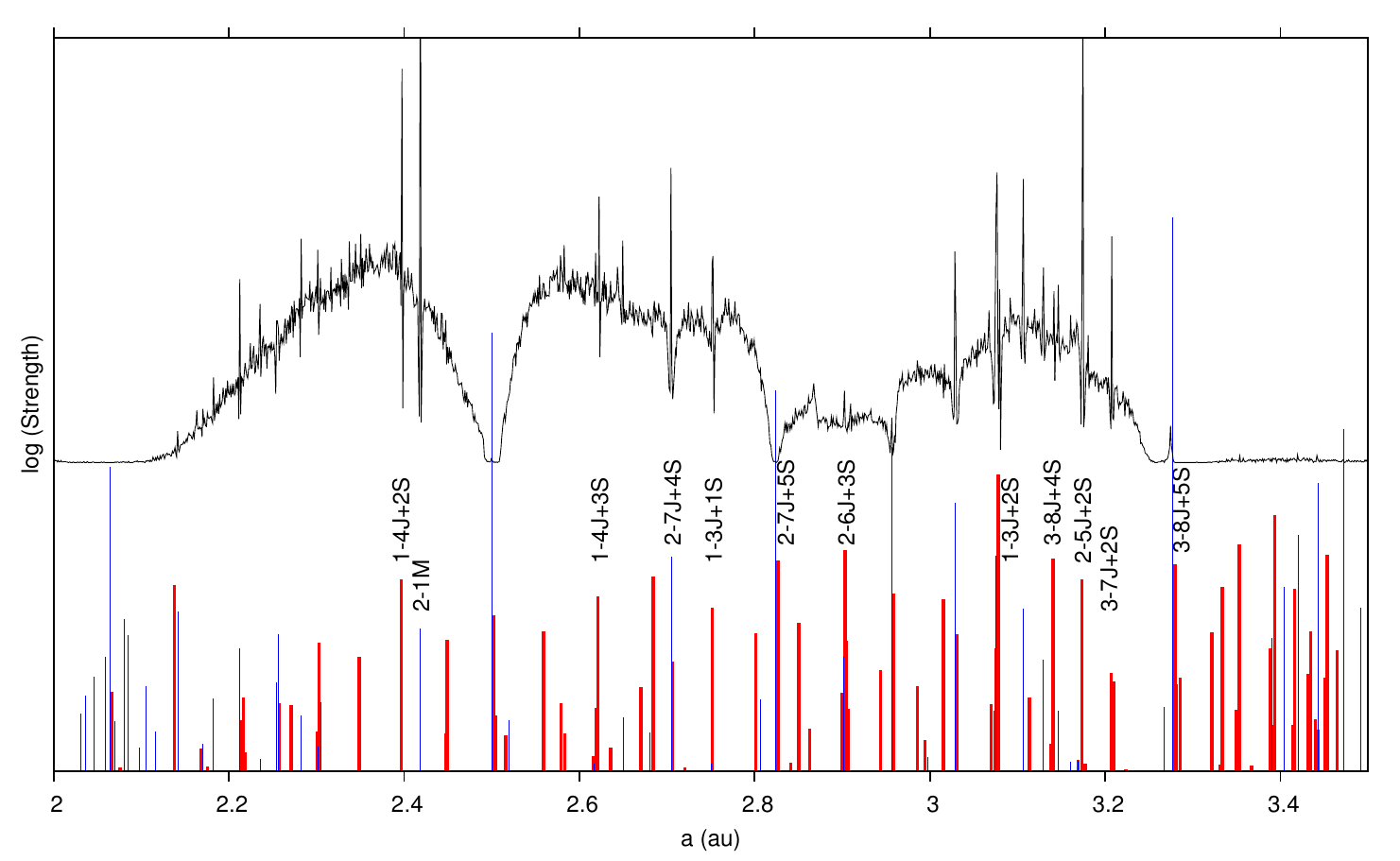}}
	\caption{Histogram of proper $a$ (black line) taken from AstDyS (hamilton.dm.unipi.it/astdys) in a normalized scale plus 2BRs (thin blue lines) and 3BRs (thick red lines). The height associated to each resonance is in logarithmic scale and indicate the relative strength calculated for a test particle with $e=0.2$, $i=10^{\circ}$ and $\omega=60^{\circ}$. The scales for 2BRs and 3BRs are different. Reproduced from \citet{2014Icar..231..273G}.}
	\label{belt}
\end{figure}

\begin{figure}[h]
	\resizebox{12cm}{!}{\includegraphics{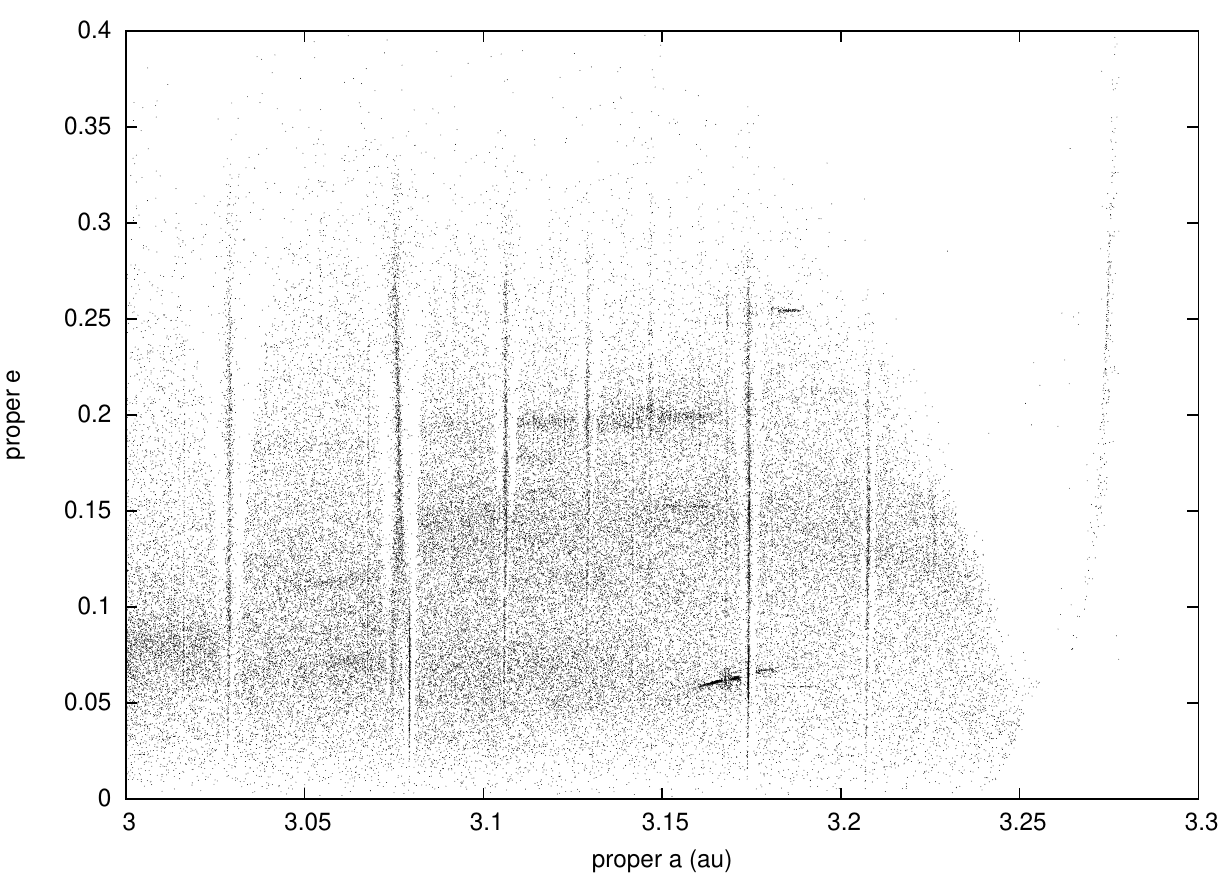}}
	\caption{Proper eccentricity versus proper semimajor axis taken from AstDyS. Resonances appear as vertical structures with increasing width for increasing $e$. The structure at $a\sim 3.075$ au is produced by a superposition of two resonances (see figure \ref{9864}) and the border at the right is due to the 2:1 resonance with Jupiter. }
	\label{proper}
\end{figure}

Resonances constitute a very rich  dynamical problem that have captivated the astronomers contributing greatly to understanding the dynamics of planetary systems. A good review of the state of arts up to the beginning of 21st century can be found for example in \citet{2002aste.conf..379N}. 
The main advances in our knowledge of MMRs of the last years can be summarized in: a) the introduction of semi-analytical  theories where the resonant disturbing function does not rely anymore on power series developments, b) a new terminology characterizing the resonances like stickiness, strength and mobility time, c) the generalization of existing models to large-inclination cases, d) the systematic study of the secular  effects inside the resonances in the transneptunian region and  e) the inclusion of other planets than Jupiter or Neptune into the scene and the study of three-body resonances.
These points are detailed respectively in Sections 2 to 6.

It is known that constructing an histogram of the asteroid's osculating $a$ the Kirkwood gaps show up. But an histogram of the \textit{proper} $a$ \citep{1999ssd..book.....M,2002aste.book..603K}  reveals a very rich structure in the main belt of asteroids due to the dynamical effects of the MMRs. Proper $a$ are very close to mean $a$ for resonant objects and consequently they are concentrated close to the nominal $a$ of the resonance generating the observed concentrations. In the histogram showed in figure \ref{belt} we can find the classic gaps but also a succession of minor gaps and evident peaks associated to weak 2BRs and 3BRs mainly involving Mars, Jupiter and Saturn. These peaks and gaps are produced by temporary captures in MMRs. 
In a plot of proper $e$ versus proper $a$ the resonant structure  is clearly shown for every resonance capable of producing some dynamical effect (figure \ref{proper}). In this figure we can see that each resonance generates a peak at the nominal position of the resonance and small gaps at both sides.

\section{Semianalytical perturbing function}
\label{semia}

The dynamical behavior of a resonant asteroid is determined by the resonant disturbing function, $R(\sigma)$, which in a first approximation can be thought as the mean over some time interval of the gravitational perturbation of the planet on the asteroid.
The disturbing function can be obtained by means of series expansions of the orbital elements followed by analytical averaging methods \citep{2007ASSL..345.....F}. Such an explicit expression for  $R(\sigma)$ is very convenient for understanding  the resonant dynamics, but as every series expansion, it is limited to restricted regions of the space of orbital elements. 
On the other hand, the increasing power of computers now allows to compute the disturbing function numerically, without using any series expansion. Even if the result is not as easy to use as an explicit expression, it is not restricted to particular domains. It can thus be used to obtain the equilibrium points and strengths of arbitrary resonances for minor bodies with arbitrary orbital elements \citep{2006Icar..184...29G, 2007arXiv0708.2080G}.

The numerical computation of a disturbing function gives place to the so-called semianalytical methods, (see for example \citet{1992A&A...257..315B,1996CeMDA..64..209T}). In these methods  the system of differential equations of the analytical methods are preserved but whenever the disturbing function appears its calculation is done numerically. In the case of a resonant problem, the numerical calculation of  $R(\sigma)$ allows us to obtain a good description of the resonance.

\begin{figure}[h]
	\resizebox{12cm}{!}{\includegraphics{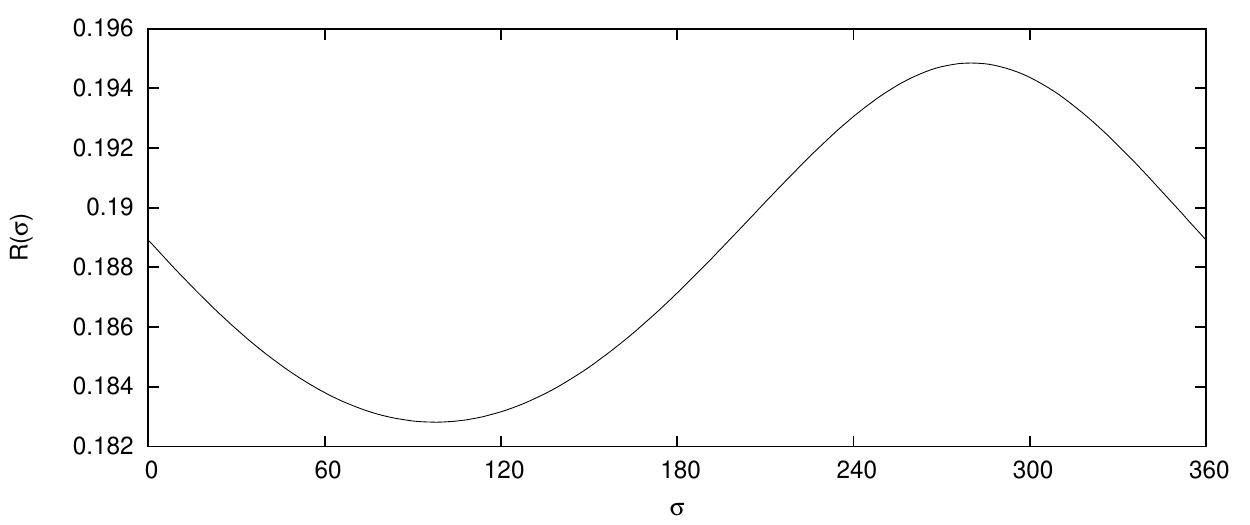}}
	\caption{Resonant disturbing function $R(\sigma)$  for resonance 3:1 with Jupiter computed numerically for an orbit with $e=0.2$,  $i=70^{\circ}$ and $\omega = 45^{\circ}$. There is a stable equilibrium point at $\sigma \sim 100^{\circ}$ (minimum  $R(\sigma)$) and an unstable one at  $\sigma \sim 280^{\circ}$ (maximum  $R(\sigma)$).}
	\label{incli}
\end{figure}

That technique shows, for example, 
that the equilibrium points for high-inclination orbits are considerably shifted from the classic values  $\sigma=0^{\circ}$ or $\sigma=180^{\circ}$ obtained for near-zero inclinations
 orbits (figure \ref{incli}). The picture of the resonant motion we have constructed  for coplanar orbits offers now a wide diversity. 
Codes for the numerical computation of the resonant disturbing function can be found at www.fisica.edu.uy/$\sim$gallardo/atlas/.

\begin{figure}[h]
	\resizebox{12cm}{!}{\includegraphics{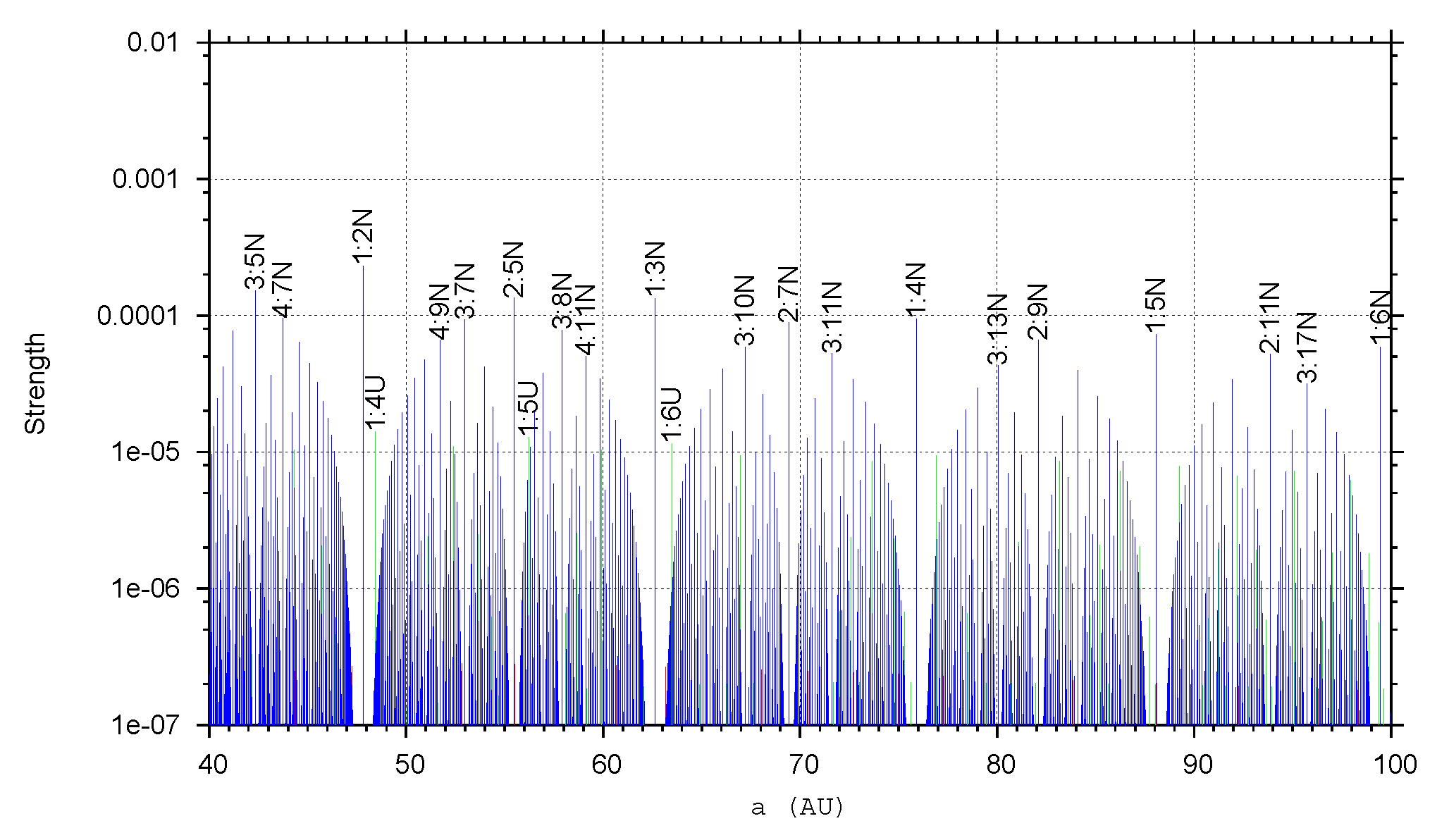}}
	\caption{System of exterior resonances with Neptune (N) and Uranus (U). The resonance's strength is calculated assuming test particles with perihelion at 32 au, $i=20^{\circ}$ and $\omega=60^{\circ}$.
		Resonances 1:k are the strongest most isolated ones and the most likely to house TNOs. Reproduced from \citet{2006Icar..184...29G}.}
	\label{40to100}
\end{figure}

\citet{2006Icar..184...29G} proposed to define $SR(e,i)=<R>-R_{min}$, more or less the semiamplitude  of the numerically computed $R(\sigma)$, as proxy for the resonance's strength and using this criteria an atlas of 2BRs in the Solar System can be constructed easily. A very complex forest of 2BRs involving the terrestrial planets emerges for asteroids with $a<2$ au and a beautiful regular pattern due to the exterior resonances with Neptune appears for $a>30$ au (figure \ref{40to100}). This figure shows that beyond Neptune the resonances of the type 1:k and 2:k dominate because of their strength but also because they are isolated from other perturbing resonances.
The known population of  TNOs in exterior MMRs with Neptune has been increased with confirmed identifications  up to the exterior 2:9 resonance at $a\sim 82$ au \citep{2016AJ....152...23V,2016AJ....152..212B}.

\section{Linking strength with width, stickiness and average time lead/lag}
\label{wssmt}

The maximum amplitude of the librations of the critical angle and the semimajor axis define an interval in au where the resonance dominates.
This can be seen for example in figure 1 of \citet{2002aste.conf..379N}.
The \textit{width} of a MMR is the extension in au covered by the resonance around its nominal position in semimajor axis defined by $n_0$. For low-inclination orbits it goes approximately with $e^q$ and consequently vanishes for near zero eccentricity orbits. Inside this region the critical angle $\sigma$ oscillates and outside these limits it circulates being the limiting curve  the separatrix \citep{2002aste.conf..379N}. 
When considering a single perturbing term, it is possible to correlate the strength $SR$ with the resonance's width as has been done by  \citet{2011MNRAS.414.1059S}
in the framework of a study of meteoroid streams.

The \textit{stickiness} can be defined as the ability of the resonance to retain minor bodies evolving inside or outside the resonance domains but sticked to the borders. In these situations the critical angle shows circulations or very large amplitude oscillations but the dynamical effects of the resonance are present and they do not allow the semimajor axis to evolve from the proximities of the resonance.
It is a  chaotic evolution which makes the semimajor axis to evolve 
switching between both sides of the resonance,
sometimes  at the right  and sometimes at the
left of the resonance. An asteroid experiencing a continuous variation  $\Delta a$ in its semimajor axis due to Yarkovsky's effect \citep{2006AREPS..34..157B}, when reaching a resonance, will be retained in a sticked motion during an interval of time that is related to the resonance's strength. This was first studied systematically by \citet{2007Icar..192..238L} in the transneptunian region obtaining a very good correspondence between the resonance's strength and the time evolving in sticking.

The \textit{average time lead/lag}, $<dtr>$, was introduced by \citet{2016ApJ...816L..31M} in the context of a study of the orbital evolution of asteroids under the Yarkovsky effect. They observed that an asteroid with a varying $a$, due for example to Yarkovsky's effect and regardless the sign of $\Delta a$, when encountering a resonance is either retained or expelled from the resonance in a time scale related to the resonance's strength. Then, the resonance's strength not only can slow down the time variations of the asteroid semimajor axis by means of sticking but also can accelerate it jumping the resonance.

\section{High-inclination orbits}
\label{space}

\begin{figure}[h]
	\resizebox{12cm}{!}{\includegraphics{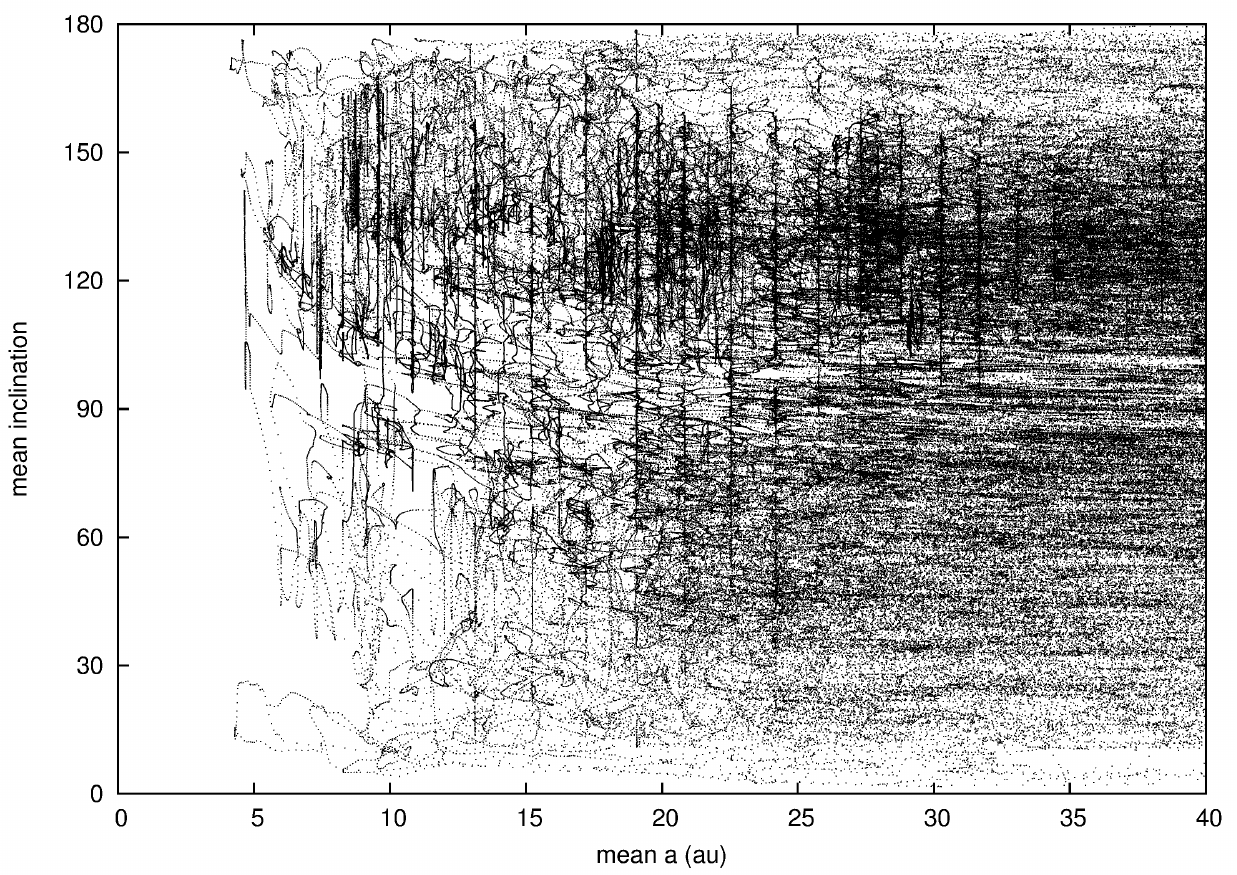}}
	\caption{Orbital states of long period comets plus clones given by their mean semimajor axis and mean inclinations calculated using time intervals of $10^4$ years integrated numerically for 4 Myrs. The vertical structures correspond to orbital states of objects captured in exterior resonances with Jupiter, mainly 1:k and 2:k. Figure adapted from \citet{2016MNRAS.461.3075F}.}
	\label{rayerio}
\end{figure}

The publication of new analytical expansions for non planar resonant problems (for example \citet{1998A&A...329..339R,2000Icar..147..129E}) stimulated the analytical studies of the resonant motion for orbits with large mutual inclination and, very recently, the problem of the extreme cases of planar retrograde and polar resonances \citep{2013CeMDA.117..405M,2017MNRAS.471.2097N}. Surprisingly, an object with almost planar retrograde orbit in resonance 1:1 with Jupiter was discovered following very closely the predictions \citep{2017Natur.543..687W}.
Moreover, numerical experiments show that captures in retrograde resonances with Jupiter and Saturn are very common for fictitious test particles \citep{2013MNRAS.436L..30M,2015MNRAS.446.1998N}.
In particular it was showed  that   
the capture of long period comets in exterior resonances of the type 1:k with Jupiter is very frequent, up to
 the resonance 1:21
\citep{2016MNRAS.461.3075F}. Figure \ref{rayerio} represents the orbital states of the population of long period comets with perihelion $<2.5$ au plus clones integrated numerically for 4 Myrs. The concentration of orbital states around some defined values of $<a>$ are due to the capture in exterior resonances with Jupiter and they occur more frequently for retrograde orbits. This highlights the fact  that the resonances are not limited to  low-inclination orbits and, on the contrary, are very relevant in orbits with large inclinations. 
Indeed, these studies show that retrograde resonant orbits are frequent for
 comets and centaurs evolving between the giant planets \citep{2017fernandez}.

\begin{figure}[h]
	\resizebox{12cm}{!}{\includegraphics{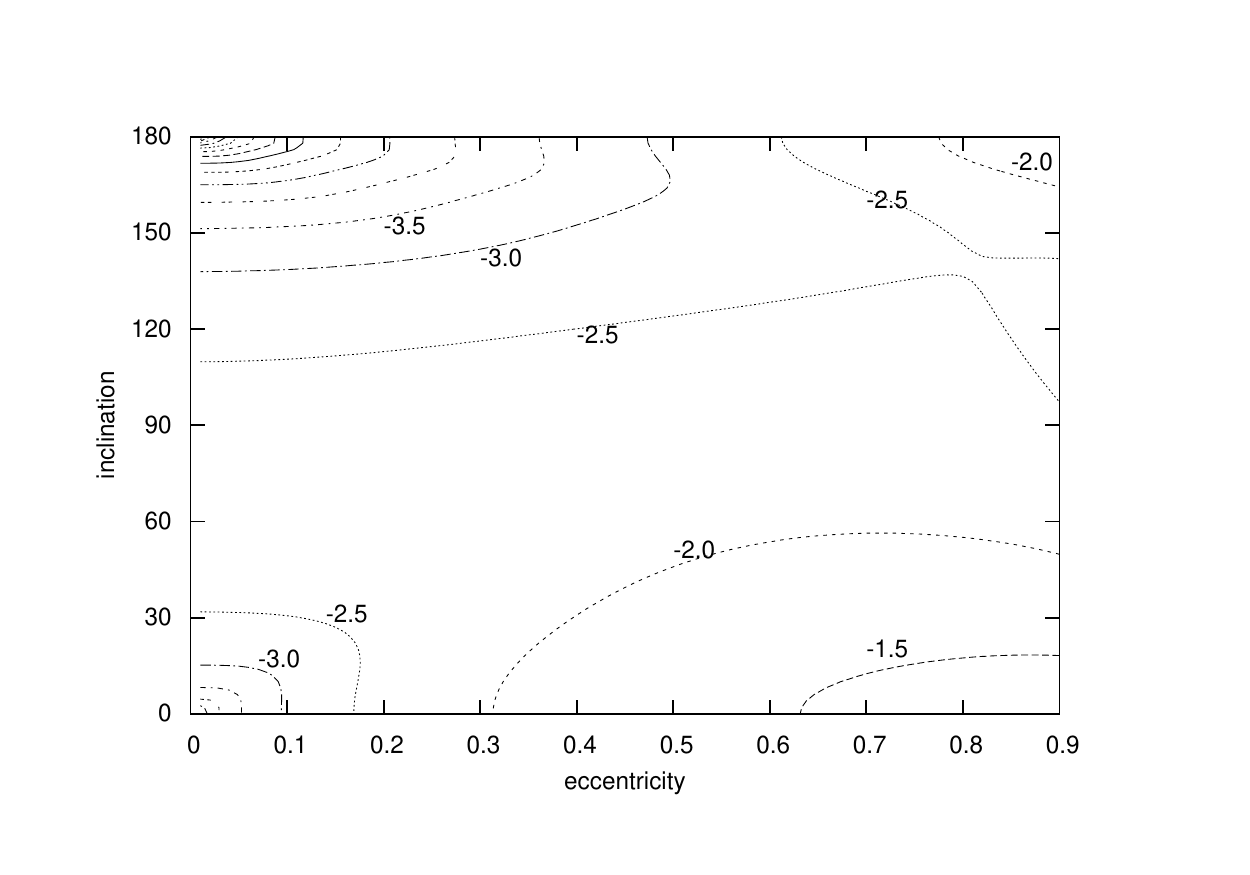}}
	\caption{Strength $SR(e,i)$ in relative units for resonance 3:1 with Jupiter. Level curves in logarithmic scale increasing from left to right. For the calculations it was taken $\omega = 45^{\circ}$. There is a strong dependence with $e$ for $i\sim 0^{\circ}$ and $i\sim 180^{\circ}$ but
		almost null dependence with $e$ for  $60^{\circ}\lesssim i \lesssim 100^{\circ}$. }
	\label{srei}
\end{figure}

The resonance's strength $SR(e,i)$ for arbitrary inclinations can be explored  following the numerical approach proposed by \citet{2006Icar..184...29G}. For example figure \ref{srei} shows  $SR(e,i)$ 
for the 3:1 resonance with Jupiter.
We can verify that for low-inclination orbits $SR \propto e^q$ as planar theories predicted, but for large inclinations this behavior is not verified anymore, and in particular, for $60^{\circ}\lesssim i \lesssim 100^{\circ}$ the strength is almost independent of the eccentricity. Moreover, for $i \sim 180^{\circ}$ it is remarkable that the dependence of $SR$ with the eccentricity is steeper than $e^q$, as was demonstrated for the coplanar retrograde resonances by \citet{2012MNRAS.424...52M}. 

\section{Long term evolution}
\label{long}

\begin{figure}[h]
	\resizebox{12cm}{!}{\includegraphics{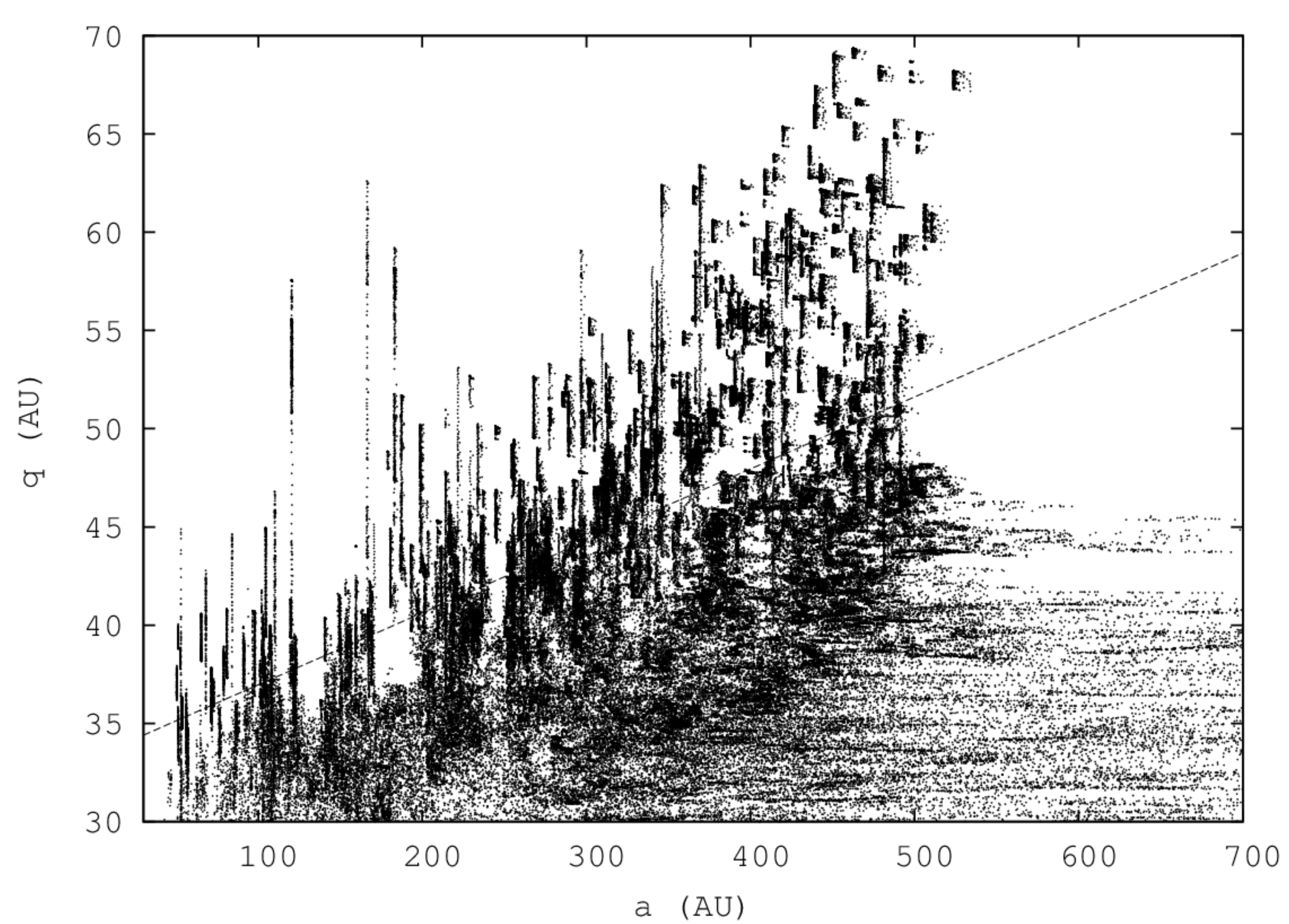}}
	\caption{Orbital states of fictitious TNOs evolving by 1 Gyr with initial inclinations between 20 and 70 degrees. Low perihelion objects experience a diffusion in $a$ until getting trapped in MMRs with Neptune and then they start to increase their perihelion distances due to KL mechanism generating the vertical structures. Reproduced from \citet{2012Icar..220..392G}.}	\label{detached}
\end{figure}

The long term secular evolution of asteroids inside 2BRs with Jupiter was very well documented in the past starting from the pioneer works by \citet{1982AJ.....87..577W} and \citet{1985Icar...63..272W}. 
 In the last years a good progress has been done in the theory of the long term evolution of TNOs in resonance with Neptune (see for example \cite{2005CeMDA..93..167S}).
There were reported large eccentricity and inclination changes associated with the Kozai-Lidov mechanism inside the MMRs \citep{2005CeMDA..91..109G,2012Icar..220..392G,2017CeMDA.127..477S,2017A&A...603A..79S}. The exterior resonances with Neptune by themselves cannot produce more than a small oscillation in the eccentricity in timescales of $10^4$ years but once the TNO is captured in resonance very often the KL mechanism appears driving large orbital changes in very long timescales ($10^8$ yrs).
Figure \ref{detached} shows the orbital states from a numerical integration of fictitious particles with initial perihelia close to Neptune that temporarily decoupled from the planet rising their perihelia due to the simultaneous action of 1:k exterior resonances with Neptune and the KL mechanism. 
It was also showed that, except for some particular inclinations around $63$ or $117$ degrees, non resonant orbits cannot experience large orbital changes.
On the contrary, large orbital changes are very common in the case
of resonant ones \citep{2012Icar..220..392G,2016CeMDA.tmp...31S,2017CeMDA.127..477S}. Large eccentricity TNOs are easily captured in 2BRs with Neptune because the resonance's strength is large for large eccentricities. Once they are trapped in a 2BR the KL mechanism may appear driving the eccentricity to oscillate slowly and with large amplitude and eventually diminishing to a value small enough to break the resonant motion due to the drop in the resonance's strength or due to the change in the topology  of the resonance \citep{2005CeMDA..91..109G,2014A&A...564A..44B,2017CeMDA.127..477S}.
In consequence, the object remains  in the transneptunian region in an hibernation state with very large perihelion distance, avoiding close encounters with Neptune. This is the mechanism that was proposed to explain the existence of eccentric large perihelion TNOs decoupled from Neptune \citep{2005CeMDA..91..109G}, leaving aside theories invoking an unknown planet. In particular \citet{2017CeMDA.127..477S}
showed that almost all known distant TNOs are located near 2BRs with Neptune in a configuration that could lead to large orbital variations due to the KL mechanism with the exceptions of 2012 VP113 and Sedna.

\section{Three-body resonances and more planets come into play}
\label{tbrs}

While Jupiter's resonant population was very well documented in the past, more recently considerable progress has been done in the study of Neptune's resonant population \citep{2012AJ....144...23G,2016AJ....152...23V}.
Although along the years there were found individual asteroids in resonance with some terrestrial planets, only
recently a large population of asteroids was identified in resonance with Mars \citep{2007Icar..190..280G,2011Icar..214..632G}, as well as some
in resonance with Venus, Earth, Saturn and Uranus \citep{2006Icar..184...29G,2008Icar..194..789C,2013A&A...551A.114D} and tens of asteroids in coorbital motion with  Ceres and Vesta  \citep{2012Icar..217...27C}.
Ceres and Vesta also imprint a secular dynamics on some asteroids as showed by \citet{2016Icar..280..300T}. 

Just after the study of resonances generated by  massive bodies other than Jupiter or Neptune started another type of MMR gained attention: resonances involving two perturbing planets, referred to as 3BRs in this paper.
When trying to obtain the resonant disturbing function for an asteroid in a 3BR with two planets we will face a very complicated problem.
So complicated that analytical expressions have been obtained  for the planar case only \citep{1999CeMDA..71..243N}. In the case of the 2BRs the disturbing function can be obtained assuming fixed orbits for the two bodies, but in order to obtain the disturbing function of a 3BR it is necessary to take into account the mutual perturbations between the three bodies. In spite of the weaker nature of the
3BRs they are more dense originating a large amount of asteroids evolving in 3BRs
mainly with Jupiter and Saturn \citep{2013Icar..222..220S,SMIRNOV2017}. 
The traces of 3BR were even investigated in meteoroid streams \citet{2016MNRAS.460.1417S}.

\begin{figure}[h]
	\resizebox{12cm}{!}{\includegraphics{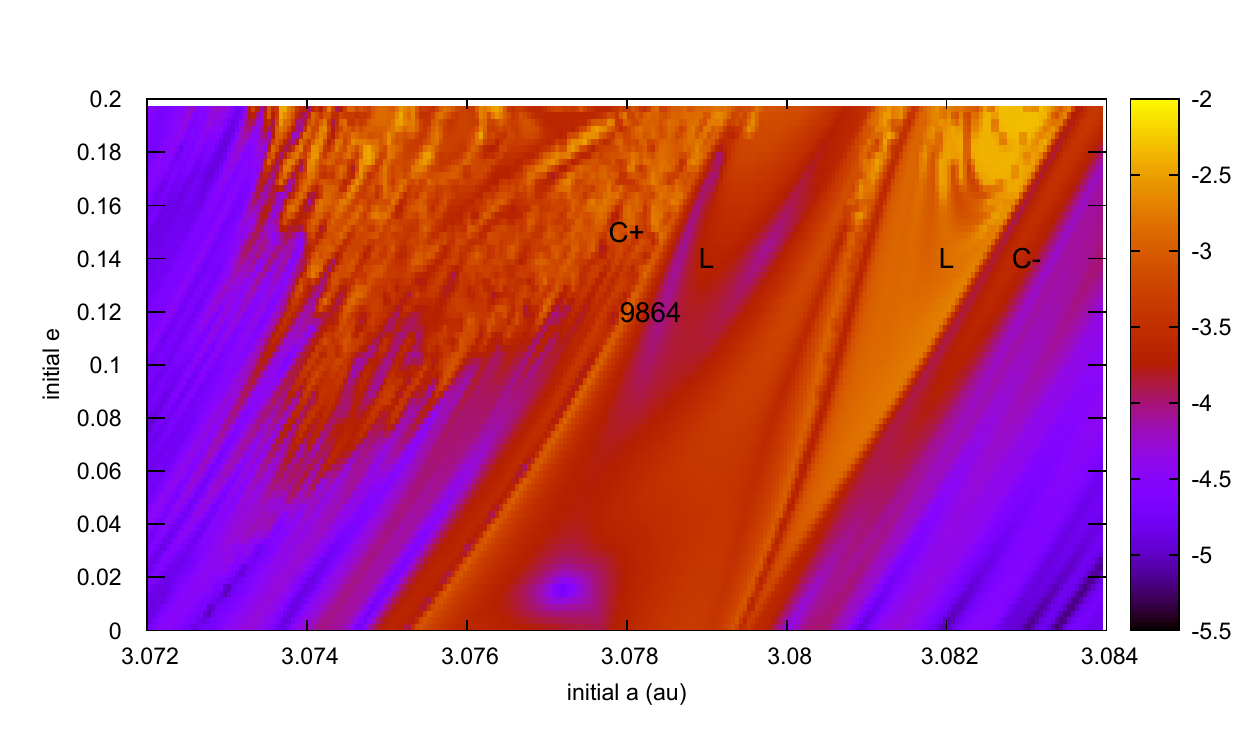}}
	\caption{Dynamical map where the color scale represents variations in $<a>$ in logarithmic scale of base 10 obtained from numerical integrations of fictitious particles in the real planetary system with initial conditions taken uniformly from a grid in $(a,e)$. Each $<a>$ is obtained by means of a moving window of 1600 yrs (smaller than the libration period) and the plotted $\Delta <a>$ correspond to the observed variations in a time span of 25000 yrs covering some libration periods. The structure at the left is the six order 2BR 11:5 with Jupiter and the one at the right is the zero order 3BR  1 - 3J + 2S showing a width of $\sim 0.004$ au.  L indicates libration of the critical angle $\sigma = \lambda_0 -3\lambda_J +2\lambda_S$ and C+ and C- indicate increasing and decreasing circulation respectively.
		The left border of this 3BR can be approximately defined by a straight line from (3.0753, 0) to (3.0794, 0.2) and the right border 	 by a straight line from (3.0795, 0) to (3.0835, 0.2).
		The approximate location of the asteroid 9864 is showed. The initial inclination is $1.5$ degrees corresponding to the orbital inclination of the same asteroid.}
	\label{9864}
\end{figure}

\begin{figure}[h]
	\resizebox{12cm}{!}{\includegraphics{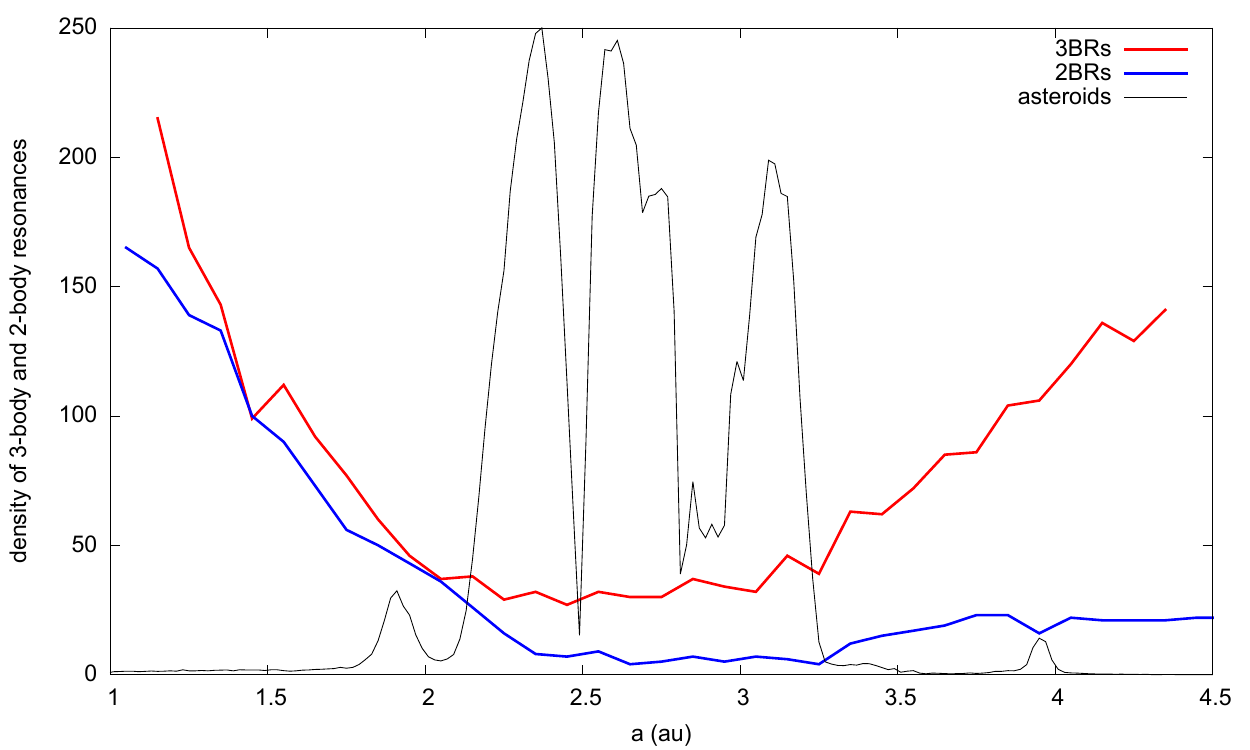}}
	\caption{Density of the strongest 2BRs and 3BRs  per 0.1 au superimposed with an histogram of the osculating semimajor axes of the asteroids (not in the same scale). For the computation of the strengths of the 2BRs it was assumed a test body with $e=0.15$, $i=5^{\circ}$ and $\omega=60^{\circ}$ and were considered all resonances with all planets up to order 30. For the computation of the strengths of the 3BRs it was assumed a test body with $e=0.2$, $i=10^{\circ}$ and $\omega=60^{\circ}$ and were considered all resonances with all planets up to order 20.}
	\label{dens}
\end{figure}

In order to obtain a practical tool for studying the 3BRs in the space of all orbital parameters \citet{2014Icar..231..273G} proposed a semianalytical method to estimate the resonant disturbing function $R(\sigma)$ and the resonance's strength. In analogy to 2BRs, following this method, it was possible to construct an atlas of 3BRs in the Solar System.
The 3BRs involving Jupiter and Saturn account for several peaks of the histogram of proper $a$ in figure \ref{belt} which are not related to 2BRs \citep{2014Icar..231..273G}.
Even if the 3BRs are very weak, the dynamics they can produce are very rich and still poorly understood. It is interesting to note that there exist zero order 3BRs, that means  $q=|k_0+k_1+k_2|=0$, a situation that for the 2BRs is restricted only to the coorbital motion, that means, the resonance 1:1. 
Zero order 3BRs have resonance strength independent of the orbital eccentricity. This has a profound significance: zero eccentricity coplanar orbits could be more affected by 3BRs than by 2BRs because 2BRs for zero eccentricity coplanar orbits have zero strength. 
In the case of the Solar System, it can be found that near a 3BR that involves Jupiter and Saturn, a high-order 2BR with Jupiter or Saturn may also be present as shown in the figure \ref{9864}. This can happen because the giant planets are close to mutual MMRs.
Figure  \ref{9864} shows a dynamical map, representing variations in mean $a$, of  a region in the main belt showing two near MMRs that can be identified in figure \ref{proper}: the 2BR 11:5 with Jupiter and the 3BR 1-3J+2S involving Jupiter and Saturn. The corresponding critical angle for this 3BR is 
$\sigma = \lambda_0 -3\lambda_J +2\lambda_S$.
We can verify in the figure that the width or strength of this 3BR is independent of the eccentricity because it is of zero order,  while the corresponding width or strength for the 2BR vanishes with $e$
because it is a high order resonance. Only for illustration, the position of the asteroid 9864 evolving in the resonance
1-3J+2S is  showed in the map.
An extensive survey of asteroids and TNOs in 3BRs with the planets in the Solar System was recently done by \citet{SMIRNOV2017}. 
According to this work the most populated 3BRs in the asteroid belt are 
2- 5J +2S, 1-3J+2S, 1-4J +2S and 1-3J +1S, each one with more than a thousand asteroids, and  in the transneptunian region the zero order resonance 2 +1U-3N houses 3 objects.
 The interaction between 2BRs and 3BRs is a complex problem that has not been studied yet but that certainly leads to chaotic diffusion. In fact figure \ref{dens} shows 
that the long lived asteroid belt is located inside a region with the lowest density of MMRs.

\section{Conclusion}
\label{conc}

Capture in resonance is a common orbital state that gives some stability, at least temporarily, to minor bodies experiencing orbital diffusion between the planets. In this period, numerical simulations and semi-analytical methods showed the relevance of the exterior resonances of the type 1:k and 2:k in particular with Jupiter and Neptune. It was found that other planets also imprint resonant signatures on asteroids and TNOs and the subtle presence of the 3BRs was exposed.
 Properties such as width, strength, stickiness and time lead/lag have been characterized, finding that all them are correlated. Semi-analytical methods have been successful in describing the long term dynamics inside MMRs and
studies devoted to high-inclination resonant orbits showed a very different panorama with respect to the planar case.
Details of the process of capture in resonance depends on the resonance's
strength, the other perturbations affecting the minor body and the
topology of the resonances in the spatial case. 
The characterization of the full three dimensional structure in the space $(a,e,i)$ of the MMRs and the study of the process of capture in  that space 
are some of the next challenges. Surely, in order to achieve these objectives, advances in the knowledge of the dynamics of resonant extrasolar systems will also play an important role.

\textbf{Acknowledgments.}
To the SOC of ACM2017 for the invitation to present this work and to the anonymous referees that contributed substantially to improve the original manuscript.
I acknowledge support from the Comisi\'on Sectorial de Investigaci\'on Cient\'ifica
(CSIC) of the University of the Republic through the project CSIC Grupo I+D
831725 - Planetary Sciences.

\bibliographystyle{elsarticle-harv}
\bibliography{biblioacm}

\end{document}